# Comparative Study of Structural and Electronic Properties of Cu-based Multinary Semiconductors

Yubo Zhang[1,2], Xun Yuan[1], Xiudong Sun[2*], Bi-Ching Shih[3], Peihong Zhang[3*], and Wenqing Zhang[1*]

[1] *State Key Laboratory of High Performance Ceramics and Superfine Microstructures, Shanghai Institute of Ceramics, Chinese Academy of Sciences, Shanghai 200050, China*

[2] *Department of Physics, Harbin Institute of Technology, Harbin 150001, China*

[3] *Department of Physics, University at Buffalo, SUNY, Buffalo, New York 14260, USA*

**ABSTRACT:**

We present a systematic and comparative study of the structural and electronic properties of Cu-based ternary and quaternary semiconductors using first-principles electronic structure approaches. The important role that Cu $d$ electrons play in determining their properties is illustrated by comparing results calculated with different exchange correlation energy functionals. We show that systematic improvement of the calculated anion displacement can be achieved by using the Heyd-Scuseria-Ernzerhof (HSE06) functional compared with the Perdew-Burke-Ernzerhof (PBE) functional. Quasiparticle band structures are then calculated within the $G^0W^0$ approximation using the crystal structures optimized within the HSE06 functional and starting from the PBE+$U$ mean-field solution. Both the calculated quasiparticle band gaps and their systematic variation with chemical constituents agree very well with experiments. We also predict that the quasiparticle band gaps of the prototypical semiconductor $Cu_2ZnSnS_4$ in the kesterite (KS) phase is 1.65 eV and that of the stannite (ST) phase is 1.40 eV. These results are also consistent with available experimental values which vary from 1.45 to 1.6 eV.

Correspondence authors: W. Zhang (wqzhang@mail.sic.ac.cn), X. Sun (xdsun@hit.edu.cn), and P. Zhang (pzhang3@buffalo.edu).

# 1. INTRODUCTION

Adamantine Cu-based ternary semiconductors (Cu-III-VI$_2$, III=Al,Ga,In; VI=S,Se,Te) and their solid solutions (CuIn$_x$Ga$_{1-x}$Se$_2$ or CIGS) have been intensively studied owing to their desired optical properties for photovoltaic application.[1-6] In recent years, Cu-based quaternary semiconductors (Cu$_2$-II-IV-VI$_4$, II=Zn,Cd; IV=Ge,Sn; VI=S,Se), especially Cu$_2$ZnSnS$_4$ and Cu$_2$ZnSnSe$_4$, have emerged as promising nontoxic, low-cost, and high efficiency materials for thin film solar cell applications.[7-11] Although it is widely recognized that these multinary semiconductors provide ample opportunities for materials design and device applications, it is still very difficult to prepare high quality ternary semiconductors and even so for quaternary semiconductors in experiments. This leads to substantial uncertainties in determining their crystal structures, electronic, and optical properties. For example, the measured anion displacement $\mu$ (defined in section 3) for CuInSe$_2$ varies from 0.2199 to 0.2499 and that for CuGaSe$_2$ varies from 0.2423 to 0.2590.[12] Even the fundamental energy gap, which is one of the most important properties for photovoltaic applications, of some of these semiconductors has not been accurately determined.

On the theory side, despite much earlier effort,[13-18] our understanding of their basic crystal structures, electronic, and optical properties is still far from satisfactory. While the tetragonal distortion $\eta \equiv c/2a \neq 1$ of these semiconductors calculated within the local density approximation (LDA) or the generalized gradient approximation (GGA) generally agrees well with experiments, both the LDA and the GGA underestimate the anion displacement parameter $\mu$ compared with reliable experiments.[12, 19] This is rather unexpected since both the LDA and the GGA have been shown to be able to describe rather well structural properties of weakly to moderately correlated materials. More intriguingly, the calculated electronic structures near the band edge[16, 19-21] and optical properties[22] are very sensitive to the anion displacement $\mu$. As a result, the uncertainty in $\mu$ (both from theory and experiment) seriously hinders our understanding of their intrinsic electronic and optical properties. Furthermore, the physics behind this strong correlation between structural properties (in this case the anion displacement $\mu$) and electronic properties of these materials has not been well understood.

The aforementioned unresolved issues have led us to investigate an important aspect of these materials that has thus far received relatively less attention, namely, the presence of strongly localized $d$ electrons and their effects on the structural and electronic properties of these materials. It is now well documented that the LDA (or GGA) cannot adequately describe the exchange-correlation ($xc$) effects of strongly localized $d$ electrons, even when the $d$ shell is fully occupied. An inadequate treatment of the $xc$ effects for the $d$ electrons within the LDA (or GGA) will certainly lead to an inaccurate account of the *pd* hybridization and the chemical bonding in these materials. Among various schemes that aim for a better treatment of strongly localized *d*-electron systems at a mean-field level, the LDA+$U$ method[23] (PBE+$U$ in this work) has been widely recognized as a simple yet powerful approach. More recently, several hybrid functionals[24, 25] aiming for a better account of the exchange interaction of localized electrons have been proposed and have been successfully applied to the study of structural and electronic properties of various systems involving localized electrons.[19, 26] In this paper, we present a systematic and comparative study of the structural and electronic properties of Cu-based ternary and quaternary semiconductors. In order to illustrate the role the Cu *d* electrons playing in determining their structural and electronic properties, we carry out calculations with different levels of approximations and exchange correlation energy functionals. We then proceed to calculate the quasiparticle band structures within the $G^0W^0$ approximation ($GW$A).[27, 28]

## 2. COMPUTATIONAL DETAILS

The calculations were carried out using the plane-wave PAW method[29, 30] as implemented in the VASP code.[31, 32] The plane wave energy cutoff is set at 400 eV and we use both the Perdew-Burke-Ernzerhof (PBE)[33] and the Heyd-Scuseria-Ernzerhof (HSE06)[24, 25] energy functionals as discussed below. For the HSE06 functional, we use a screening parameter $\omega$ of 0.2 bohr$^{-1}$ and a mixing parameter $\alpha$ of 0.25. The Brillouin-zone integration is carried out with a 4×4×4 Monkhorst-Pack *k* mesh for the crystal structure optimization and a 8×8×8 *k* mesh for the $G^0W^0$ calculations. The kinetic energy cutoff for the dielectric screening in the $G^0W^0$ calculation is set at 186 eV. We have tested the convergence of the $G^0W^0$ calculation and the calculated band gaps shall converge within 0.05 eV using about 200 unoccupied bands.

For the PBE+$U$ calculations, since different materials may have rather different dielectric screening behaviors, it is very important to calculate the screened on-site Coulomb energy ($U$) from first-principles. There are various empirical methods that fit (or estimate) the parameter of the screened on-site Coulomb energy ($U$). However, depending on the fitting procedure, the value of $U$ sometimes varies significantly. For example, the screened Coulomb $U$ for the semicore d electrons in ZnO used in literature varies from 4.7 eV to 13 eV.[34-38] Without a first-principles approach, this physical quantity becomes a "convenient" adjustable parameter, which sometimes causes significant confusion in the literature. We have recently implemented[39] a combined Wannier orbitals and constrained random phase approximation (cRPA) method[40] to calculate the on-site Coulomb and exchange energies. We first construct the maximally localized Wannier orbitals[41] for all valence states calculated within the PBE+$U$ method using some initial $U$ and $J$.

The screened Coulomb and exchange matrices are then calculated:

$$U_{ij} =< i,j | W_r(r,r') | i,j >, \quad J_{ij} =< i,j | W_r(r,r') | j,i >, \tag{1}$$

where $W_r(r,r')$ is the effective (screened) Coulomb interaction calculated within the cRPA and $|i>$ and $|j>$ denote the Wannier orbitals for Cu $d$ states. The familiar $U$ and $J$ parameters used in the LDA+$U$ method are the averaged values of the $U$ and $J$ matrices:

$$U = \frac{1}{(2l+1)^2} \sum_{ij} U_{ij}, \quad J = \frac{1}{2l(2l+1)} \sum_{i \neq j} J_{ij} \tag{2}$$

where $l = 2$ for $d$ states. The calculation is terminated when the output $U$ and $J$ are the same as the input values. In a crystal field with a $T_d$ symmetry, the five $d$ states split into an $e$ doublet and a $t_2$ triplet. Table I shows the calculated matrix elements for both the $U$ and $J$ matrices. We obtained $U$ = 4.85 eV and $J$ = 0.82 eV for the Cu $d$ states in $CuGaS_2$. This gives an effective $U_{eff} = U - J \approx 4.0$ eV. These values are also expected to be valid for other Cu-based multinary semiconductors since they have similar structure and thus chemical bonding characteristic. It is interesting to point out that the $e$ states are more localized than the $t_2$ states as will be discussed in more detail later. The diagonal elements of the $U$ matrix in Table I shows slightly larger Coulomb energies for the $e$ states (6.21 ~ 6.23 eV) than those for the $t_2$ states (6.11 ~ 6.12 eV). This slight difference in localization is also observed in the calculated Wannier spreads. The Wannier spread for the $e$ states is about 0.60 Å$^2$, and that for $t_2$ states is about 0.73 Å$^2$.

Since we have just recently implemented this method, its validity and reliability require more careful assessment. We have carried out calculations for the on-site Coulomb and exchange energies for a wide range of materials and systems, including transition metal oxides, transition metals, and strongly localized defect states. Our results are consistent with the values used in the literature. For example, the calculated on-site bare Coulomb energies for $d$ electrons are 19.1 eV, 23.1 eV and 24.7 eV for MnO, CoO, and NiO respectively. The corresponding screened on-site Coulomb energies are 6.3 eV (MnO), 6.9 eV (CoO), and 6.7 eV (NiO). These results fall within the published values calculated with the constrained LDA approach.[23, 42] In addition, we have carried out calculations for transition metals. Our results are again consistent with published values.[41, 43] We will report these results and our numerical implementation in a follow-up publication. We would also like to mention that the correction term arising from the on-site Coulomb correlation within the PBE+$U$ (or LDA+$U$) method is appreciable only for strongly localized electrons and becomes negligible for delocalized electrons. We have calculated the inter-site screened Coulomb and exchange energies between $d$ and neighboring $sp$ electrons and found that they are several times smaller than the values for $d$ electrons. For example, the screened Coulomb $U$ between the Cu $d$ electrons and neighbor $p$ electrons is only about 1.0 eV.

## 3. BRIEF DESCRIPTION OF CRYSTAL STRUCTURES

Before we proceed to present our results, it is useful to briefly describe the structures of Cu-based ternary and quaternary semiconductors. The chalcopyrite (CH) structure [space group $I\bar{4}2d$, see Fig. 1 (a)] is the most stable phase for many of Cu-based ternary semiconductors with tetrahedral bonding (e.g., CuGaS$_2$). This structure can be obtained by cation mutation[3] of their II-VI binary analogs (e.g., ZnS). Further mutation of the group-III atoms in ternary semiconductors to II and IV atoms (e.g., Ga in CuGaS$_2$ to Zn and Sn) leads to quaternary semiconductors (e.g., Cu$_2$ZnSnS$_4$) with either the stannite (ST) [space group $I\bar{4}2m$, see Fig. 1 (b)] or the kesterite (KS) [space group $I\bar{4}$, see Fig. 1 (c)] structure. This cation-mutation strategy has been adopted by several groups to proposed novel semiconductors with desired properties.[1-3]

The building block of all the above mentioned structures is a tetrahedron consisting of a centered anion atom and four tetrahedrally bonded cation atoms (see Fig. 2). In binary systems, the group-VI anion is surrounded by four identical group-II cations. Moving from binary to ternary systems, the VI anion is surrounded by two Cu and two group-III atoms, leading to a distortion to the tetrahedron unit. This distortion can be described by a parameter called anion displacement $\mu$,[44] which is 0.25 for an ideal tetrahedron in cubic binary systems. Taking Cu-III-VI$_2$ as an example [see Fig. 1 (a)], the anion parameter is related to Cu-VI and III-VI bond lengths (denoted as $R_{\text{Cu-VI}}$ and $R_{\text{III-VI}}$, respectively) as defined by

$$\mu_{\text{CH}} = 0.25 + \left(R_{\text{Cu-VI}}^2 - R_{\text{III-VI}}^2\right)/a^2 . \qquad (3)$$

For quaternary semiconductors, there are three different bonds in the VI-centered tetrahedron in the ST structure and four in the KS structure [see Fig. 1 (b) and (c)]. This is because the cation layers alternate along the lattice $c$ direction with a sequence of Cu-Cu/II-IV/Cu-Cu/IV-II in the ST structure and of Cu-II/Cu-IV/II-Cu/IV-Cu in the KS structure. As a result, the anion displacement is defined differently:

$$\mu_{\text{ST}} = 0.25 + \left[R_{\text{Cu-VI}}^2 - \frac{\left(R_{\text{II-VI}}^2 + R_{\text{IV-VI}}^2\right)}{2}\right]\bigg/a^2 , \text{ and} \qquad (4)$$

$$\mu_{KS} = 0.25 + \left[\frac{\left(R_{(Cu\text{-}VI)-1}^2 + R_{(Cu\text{-}VI)-2}^2\right)}{2} - \frac{\left(R_{II\text{-}VI}^2 + R_{IV\text{-}VI}^2\right)}{2}\right]\Big/a^2. \tag{5}$$

The anion displacement $\mu$, which characterizes the structural distortion of the tetrahedron building blocks, critically links to the chemical bonding and influences the electronic structure near the band edge of Cu-based multinary semiconductors. Taking CH ternary semiconductors (e.g. CuGaS$_2$) as an example, the anion displacement $\mu_{CH}$ measures the difference in bond length between the Cu-VI and III-VI bonds. These bond lengths in turn reveal the chemical bonding and hybridization between relevant atomic states. Since the top valence states are mainly derived from Cu $d$ and VI $p$ orbitals, their properties closely correlate with the Cu-VI bond length. For example, a smaller VI anion and a stronger $pd$ hybridization will presumably result in a shorter Cu-VI bond, thus a smaller $\mu_{CH}$. On the other hand, the bottom of the conduction bands are mainly composed of III $s$ and VI $p$ orbitals, which are also more sensitively influenced by the III-VI bond. Of course, density functional theory (DFT) based first-principles electronic structure methods are supposedly able to determine the degree of $pd$ hybridization and the Cu-VI bond length self-consistently provided that the underlying energy functionals can adequately treat the exchange-correlation effects of the system. However, this is not guaranteed for systems involving strongly localized $d$ electrons. It has been well recognized that the LDA (or GGA) fails in many aspects when applied to systems containing strongly localized electrons. In the following section, we carefully exam the effects of different energy functionals on the structural and electronic properties of Cu-based semiconductors.

## 4. RESULTS

### 4.1. General Characteristics of Cu $d$ Electrons in Cu-based Multinary Semiconductors and Theoretical Challenges

Electronic and structural properties of Cu-based multinary semiconductors are influenced by the subtle interplay between the covalent bonding and the localization tendency of Cu $d$ electrons. On one hand, $d$ electrons in late transition metal elements are fairly localized and strongly interact among themselves with a characteristic energy, the *screened* onsite Coulomb $U$. On the other hand, these semicore $d$ electrons are relatively high (shallow) in energy and can couple strongly with VI $p$ valence states in the system. Proper treatment of these localized $d$ electrons, especially their coupling with other valence states, remains a challenging problem.

Taking the CH CuGaS$_2$ as an example, we plot in Fig. 3 the projected density of states (DOS) calculated using the PBE energy functional. Some general features are worth mentioning. First, it is obvious that Cu $d$ states spread in a wide energy range of 0 to -5 eV below the valence band maxim (VBM). Second, the five Cu $d$ states split into $e$ and $t_2$ states under the influence of the crystal field and they show very different behaviors in forming chemical bonds with neighboring atoms.[26] The $t_2$ states can hybridize with the valence $p$ states, therefore they participate strongly in chemical bonding. The bonding behavior or the $t_2$ states can also be seen from their charge distribution [shown in the insert of Fig. 3 (b)] which shows that the charge points towards the cell edges (where S atoms locate). The $e$ states, in contrast, are better characterized as nonbonding states since their hybridization with valence $p$ states is prohibited by symmetry and their charge distribution pointing towards the face center of the cell [shown in the insert of Fig. 3 (a)]. Strictly speaking, the $T_d$ symmetry is only approximate in these systems, but the above discussion is still valid. As a result, whereas the $e$ states are sharply peaked around -1.5 to -2 eV in the DOS plot, the hybridization between $t_2(d)$ states and S $p$ states forms bonding (around -3.5 to -5 eV) and anti-bonding (around 0 to -2 eV) states. The anti-bonding $pd$ states overlap in energy with nonbonding $e$ states. Both the intrinsic localization of Cu $d$ states and the covalent hybridization play important roles in determining the electronic properties of these systems. It is this dual nature of $d$ electrons in these systems that requires delicate treatments, and the conventional LDA (or GGA) may not be adequate. We mention that for quaternary semiconductors, the $d$ electrons from group-II elements, for example Zn $3d$ in $Cu_2ZnSnS_4$, are also highly localized. However, the Zn $3d$ states are much lower in energy [~ -10 eV from the VBM, see Fig. 7 (c)] and they do not strongly affect the electronic structure for Cu-based multinary semiconductors. We will come back to this point later.

The important role that the Cu $d$ electrons plays and the difficulty in treating these localized electrons pose significant challenges to our understanding of properties of Cu-based multinary semiconductors. Fortunately, the similarity in chemical bonding of these semiconductors makes it possible to carry out a systematic investigation of their electronic and structural properties. In the following, we first compare the experimentally measured anion displacements of ternary semiconductors with those calculated using the PBE and the HSE06 functionals.

**4.2. Anion Displacement of Ternary Semiconductors**

As discussed in Section 3, the anion displacement strongly correlates with the electronic structures near the band edges of Cu-based multinary semiconductors. Theoretical calculations based on the LDA (or GGA) have systematically underestimated[19] this parameter, rendering the description of the electronic structure within these approximations questionable. Unfortunately, experimentally measured values[12] also have very large uncertainties (see vertical bars in Fig. 4), possibly a result of crystalline imperfection such as partial disordering (especially cations) and the presence of other defects. As mentioned earlier, the measured $\mu$ of CuInSe$_2$ varies from 0.2199 to 0.2499 while that of CuGaSe$_2$ varies from 0.2423 to 0.2590.[12] The uncertainty (or error) in $\mu$ value leads to a significant change in the calculated band gap and the electronic structure near the band edge. For example, the calculated band gap [within the PBE+$U$ ($U$ = 4 eV) approach] of both CuInSe$_2$ and CuGaSe$_2$ can change by as much as 0.4 eV depending on the $\mu$ value (show above) used. J. Vidal *et al*. also reported that the calculated band gap depends strongly on the anion displacement $\mu$.[19] This is mainly because anion displacement, which directly related to the Cu-S and Ga-S bonding, sensitively influences the electronic properties of the VBM (composed of Cu d and S p states) and the CBM (composed of Ga s and S p states). Therefore, theory must be able to reliably predict the $\mu$ parameter in order to correctly describe the electronic structures of these materials.

Earlier theoretical results calculated within both the LDA and GGA consistently underestimated the $\mu$ parameter as a result of inadequate treatments of the localized Cu d within these approximations. For example, the calculated values[16] within in the LDA are 0.2225 and 0.2510 for CuInSe$_2$ and CuGaSe$_2$, respectively, lying at the lower end of experimental range. Our calculations within PBE, as shown in Fig. 4, also reproduce this trend. The anion displacements calculated within the PBE are generally lower than the measured values for most of systems even with the experimental uncertainty taken into consideration (see Fig. 4). Taking the bond lengths as sum of atomic covalent radii,[14] the anion displacements can also be estimated, giving the so-called *bond rule* result (see Fig. 4). This approach again gives the correct trend of the variation of the $\mu$ parameter but the values are generally too large compared with experiments.

The failure of both the LDA (or GGA) and the bond rule approach has the same physical origin, i.e., both methods cannot correctly capture the physics of the localized d electrons and therefore the degree of pd hybridization. Whereas the LDA (or GGA) tends to delocalize the d electrons and underestimates the pd hybridization (therefore the covalent bonding) behavior between the Cu d electrons and VI p electrons, the simple bond rule assumes an ideal covalent bonding scenario. It is well understood that the degree of pd hybridization depends on the relative energy levels and the overlap of wave functions. The calculated d levels are generally too shallow within the LDA (or GGA), which may result in either an overestimate or an underestimate of the pd hybridization depending of the relative positions of the d and p levels as will be discussed in more details in Section 4.3. We emphasize that it is the subtle balance between the localization and the bonding tendency of the Cu d electrons that makes theoretical treatments of these materials very difficult.

We further proceed with calculations using the HSE06 hybrid functional, a widely used functional that has been shown to better describe the exchange correlation effects of localized electrons. The HSE06 functional enhances the localization of Cu $d$ states and elongates the Cu-VI bonds, resulting in an increase of the anion displacement $\mu$. We mention that the HSE06 functional also enhances $pd$ hybridization because the energy levels of localized Cu $d$ states are pushed down and thus move closer to VI $p$ states. This effect may tend to shorten Cu-VI bonds. Overall, it seems that the localization effect dominates and the $\mu$ parameters calculated using the HSE06 functional are systematically improved (see Fig. 4). For example, the calculated $\mu$ is 0.2490 within the PBE functional and 0.2549 within the HSE06 for $CuGaS_2$, while the experimental values[12] range from 0.2500 to 0.2720. Our results agree with earlier calculations[19] using the HSE06 functional which gives a value of 0.229 for $CuInS_2$ and 0.227 for $CuInSe_2$. A more recent work[45] reported $\mu$ values of 0.2537, 0.2266, 0.2508, and 0.2259 for $CuGaS_2$, $CuInS_2$, $CuGaSe_2$, and $CuInSe_2$ within the HSE06 functional. The systematic improvement of the calculated anion displacement using the HSE06 hybrid functional compared with the LDA (or GGA) functional can be attribute to the fact that the HSE06 functional is able to better capture the short-range screened Hartree-Fock energy for localized electrons.

Unlike ternary semiconductors in which $\mu_{CH}$ uniquely defined the deformed tetrahedrons, the anion displacement, i.e., $\mu_{ST}$ or $\mu_{KS}$, alone does not provide a full picture of the local bonding structure in quaternary semiconductors because the anion displacements towards the group-II (e.g., Zn in $Cu_2ZnSnS_4$) and group-IV (e.g., Sn in $Cu_2ZnSnS_4$) atoms are generally different. We will present our results for quaternary semiconductors elsewhere.

**4.3. Electronic Properties of Multinary Semiconductors**

The above discussions have established that both the bonding and the localization nature of Cu $d$ electrons play an important role in determining the local structure and the electronic properties of Cu-based multinary semiconductors. We have shown that a systematic improvement to the calculated anion displacements is achieved using the HSE06 functional compared with the PBE functional. In this section, we compare the electronic properties of these materials calculated at different theory levels.

There are a few technical details that should be mentioned before we present our results. First, the $GW$ approximation[27, 28] is still the state-of-the-art many-body perturbation technique for calculating the quasiparticle properties of moderately correlated materials provided that a reasonable mean-field solution is available. As we have discussed in the previous sections, the PBE functional cannot treat the correlation effects of localized $d$ electrons adequately. Therefore, a straightforward $G^0W^0$ calculation based on the PBE solution may not be able to accurately predict the electronic properties of these multinary semiconductors. Recently, $G^0W^0$ calculations starting from the LDA+$U$ (or GGA+$U$) solutions have been shown to give promising results for systems containing localized $d$ or $f$ electrons.[42, 46-49] Second, although the HSE06 functional and its combination with the $G^0W^0$ approximation have attracted much attention recently, the applicability of this approach still awaits more testing and verification. As will be discussed later, the HSE06 functional seems to over-bind deeper valence states such as the S 3$s$ state in CuGaS$_2$. Even after applying the $G^0W^0$ correction to the HSE06 results, the S 3$s$ state is still too deep compared to experiment. Therefore, we will mainly focus our discussion on the PBE+$U$+$G^0W^0$ results. In the following sections, all electronic structure calculations were carried out using the structures optimized within the HSE06 functional.

### A. Quasiparticle properties of CuGaS$_2$ and Cu$_2$ZnSnS$_4$

Although we have calculated the screened Coulomb $U$ ($U_{\text{eff}}$ = $U$ - $J$ = 4.0 eV) for Cu $d$ electrons in these multinary semiconductors using a newly developed method as discussed earlier,[39] the validity of this value needs to be verified. In the following, we first validate the PBE+$U$+$G^0W^0$ approach and the value of $U$ using CuGaS$_2$ and Cu$_2$ZnSnS$_4$ as examples. Fig. 5 shows the calculated band gaps of CuGaS$_2$ and Cu$_2$ZnSnS$_4$ as a function of $U$. For CuGaS$_2$ [(see Fig. 5 (a)], the conventional PBE+$G^0W^0$ approach (corresponding to using a $U$ = 0 eV) gives a band gap of 1.68 eV which is about 0.75 eV smaller than the experimental value. This value is improved to 2.41 eV (to be compared to 2.43 eV measured experimentally[50]) when an $U_{\text{eff}}$ = 4.0 eV is used. For the quaternary semiconductor Cu$_2$ZnSnS$_4$, our calculations [(see Fig. 5 (b)] give a band gap of 1.40 eV for the ST structure and 1.65 eV for the KS structure. Experimentally, the most widely cited gap is about 1.5 eV[8, 51-56] irrespective of their phases. However, we would like to mention that the experimental values vary from 1.45 to 1.6 eV.[11, 57, 58] Interestingly, the lower end of the experimental gap agrees well with the calculated band gap for the ST structure whereas the upper end agrees with our result for the KS structure. It would be interesting if high quality single phase Cu$_2$ZnSnS$_4$ can be synthesized and our predictions can be verified. Our calculations in general agree well with the experimental values for the two prototypical systems. These results clearly support the value of $U$ used in our calculation and the applicability of the PBE+$U$+$G^0W^0$ approach in treating electronic structure of Cu-based multinary semiconductors.

We have briefly mentioned that although there are other semicore $d$ electrons (in addition to the Cu 3$d$) in quaternary semiconductors, they do not strongly affect the electronic properties near the band edge. Table II compares the band gap of the KS structure Cu$_2$ZnSnS$_4$ calculated within the PBE+$G^0W^0$ and the PBE+$U$+$G^0W^0$ approaches with an on-site $U$ applied to the $d$ electrons of different elements, i.e., Cu 3$d$, Zn 3$d$ and Sn 4$d$. It is clear that applying an on-site $U$ to the Cu 3$d$ electrons has the most significant effects on the calculated band gap.

The band gap alone does not provide a full picture of the electronic structure of a material. Here we compare the DOS of CuGaS$_2$ calculated within the PBE+$U$+$G^0W^0$ approach and the X-ray photoemission spectroscopy (XPS) data.[59] In addition to the well-known band gap problem, the PBE functional under-binds the Cu $d$ states [see Fig. 6 (a)], especially the nonbonding $e$ states (see Fig. 3 for the projected DOS). Applying an on-site Coulomb $U$ pushes Cu $d$ states down [see Fig. 6 (b)] and enhances the $pd$ hybridization. As a result, all major peaks observed in XPS are well reproduced in both the PBE+$U$ [see Fig. 6 (b)] and the PBE+$U$+$G^0W^0$ [see Fig. 6 (c)] approaches. It should be pointed out that the $G^0W^0$ correction to the PBE+$U$ quasiparticle energy is not a simple scissors shift to the band energy for all Cu-based multinary semiconductors, although for CuGaS$_2$ it does seem to act as a scissors operator.

We also include the DOS calculated within the PBE+$G^0W^0$, the HSE06, and the HSE06+$G^0W^0$ approaches for comparison. Besides the fact that the PBE+$G^0W^0$ approach still underestimates the band gap (1.68 eV, to be compared with the experimental value of 2.43 eV), it also leads features that do not seem to agree with experiment. For example, the separation between bonding and anti-bonding states (both are occupied) seems to be too large compared with experiment energy.[59] The nonbonding $d$ states are also too shallow in energy. The HSE06 functional has often been used to study the electronic structure for these semiconductors and has been shown to result in improved band structures, especially band gaps.[19, 26, 45, 60] Our results [see Fig. 6 (e)] also confirm that the HSE06 functional produces a gap of 2.14 eV (2.22 eV in Ref. 26 and 2.44 eV in Ref. [45]), which is only slightly smaller than measured value. We also notice that the HSE06 functional and the PBE+$U$+$G^0W^0$ approach give similar band structure in the energy window of 0 ~ -6 eV below the VBM. However, for the low-lying valence states such as S 3$s$ states, the HSE06 functional seems to overestimate the band energy. The band energy for S 3$s$ states calculated with the HSE06 functional is about 1 eV too deep compared with the experimental values[59, 61] and results obtained using the PBE or the PBE+$U$ approach. Our PBE results agree with previous theoretical predictions.[13] This raises an interesting question regarding the validity of using a single screening parameter for all electronic states. Apply the $G^0W^0$ correction to the HSE06 mean-field solution only gives marginal improvement in both the fundamental band gap and the position of S 3$s$ states. In addition, the separation between the bonding an anti-bonding valence states calculated within the HSE06+$G^0W^0$ approach is very similar to that calculated within the PBE+$G^0W^0$ approach, and does not seem to agree with experiment as discussed before.

The above discussions have demonstrated that the PBE+$U$+$G^0W^0$ approach is able to reproduce both the band gap and band structure of the prototypical Cu-based semiconductor CuGaS$_2$, provided that faithful crystal structures (in this case, we use the structures optimized within the HSE06 functional) and on-site Coulomb energies are used. This approach should be also valid for other Cu-based multinary semiconductors considering their chemical similarity.

**B. Quasiparticle band structures and gaps of other multinary semiconductors**

We further calculate the quasiparticle band structures of four most interesting Cu-based semiconductors for solar cell application, i.e., CH CuGaSe$_2$, CH CuInSe$_2$, KS Cu$_2$ZnSnS$_4$, and KS Cu$_2$ZnSnSe$_4$ using the PBE+$U$+$G^0W^0$ approach as shown in Fig. 7. All of these materials have direct gaps and the calculated band gaps agree well with experimental values within the theoretical accuracy. The band structures of these semiconductors show some interesting similarities. The (occupied) valence bands near the band edge (i.e., 0 ~ -6 eV from VBM) are derived from strongly hybridized Cu $d$ and VI $p$ states, whereas the conduction bands are composed of anti-bonding states of III $s$ and VI $p$ in ternary semiconductors (IV $s$ and VI $p$ in quaternary semiconductors). The bonding and anti-bonding $pd$ states are separated by an energy gap near -3 eV from the VBM. The fundamental band gap varies significantly from sulfide to selenide semiconductors (from 1.65 eV for Cu$_2$ZnSnS$_4$ to 1.08 eV for Cu$_2$ZnSnSe$_4$), which can be understood in terms of different anion electronegativities. However, this band gap change is significantly smaller than the change in their binary analogs.[14, 62] The band gap of the ternary semiconductor is also substantially larger than that of its quaternary counterpart.[63] For example, the band gaps change from 1.60 eV for CuGaSe$_2$ to 1.08 eV for Cu$_2$ZnSnSe$_4$, and from 2.43 eV for CuGaS$_2$ to 1.65 eV for Cu$_2$ZnSnS$_4$. This band gap reduction is similar to the band gap anomaly observed in ternary semiconductors as compared to their binary analogs which has been well-documented.[14] Besides the contribution from cation electronegativity, this band gap anomaly (i.e., band gap reduction from ternary to quaternary semiconductors) mainly comes from the conduction band minimum (CBM) state which is derived from an anti-bonding combination of IV $s$ and VI $p$ states[63] as discuss below using CuGaSe$_2$ and Cu$_2$ZnSnSe$_4$ as examples.

First, as mentioned earlier, the presence of Zn in CuZnSnSe$_4$ has minor effects on the band gap since the Zn 3$d$ states are located at much lower energies [see Fig. 7 (*d*)] than the Cu 3$d$ and Se 4$p$ states. Also, the Zn 4$s$ state lies at a higher energy than the CBM and therefore has no significant effects on the band gap. The CBM state is derived from the anti-bonding combination of Ga 4$s$ and Se 4$p$ in CuGaSe$_2$, and from Sn 5$s$ and Se 4$p$ in Cu$_2$ZnSnSe$_4$. Since the atomic energy level of Sn 5$s$ state are much deeper than that of Ga 4$s$ state,[63-65] the anti-bonding state of Sn 5*s* and Se 4*p* is much lower than that of Ga 4*s* and Se 4*p* [see Fig. 7 (a) and (d)]. As a result, the CBM state in CuZnSnSe$_4$ is lower in energy than in CuGaSe$_2$. Thus, as discussed by Chen *et al*,[63] the relatively low energy of the anti-bonding combination of Sn 5*s* and Se 4*p* states in CuZnSnSe$_4$ is responsible for the gap reduction in Cu$_2$ZnSnSe$_4$ compared with CuGaSe$_2$.

We now discuss the band gaps of other Cu-based ternary semiconductors. The band gaps of nine ternary semiconductors Cu-III-VI$_2$ (III=Al,Ga,In; VI=S,Se,Te) are calculated within the PBE, the PBE+$U$, and the PBE+$U$+$G^0W^0$ approaches and are compared with available experimental results as shown in Fig. 8. For systems containing high Z elements, a correction (-0.23 eV for CuGaTe$_2$ and -0.2 eV for CuInTe$_2$) from the spin-orbit coupling[66] is included in the calculated band gap. Data points fall on the diagonal line indicate a perfect agreement with experiment. As it can be seen from Fig. 8, our results for all ternary semiconductors reproduce well the systematic trend of the chemical variation of the band gaps. In addition, calculated band gap agree quantitatively with available experimental results within ±0.2 eV. For sulfide and selenide semiconductors, while the PBE+$U$ approach improves the gaps by about 0.35 ~ 0.45 eV, the $G^0W^0$ correction further opens up the gap by 0.85 ~ 1.15 eV. As to the telluride semiconductors, the PBE+$U$ approach improves the gaps by only about 0.25 eV, and again the $G^0W^0$ correction is able to bring the band gaps to their experimental values. We mention that the band gaps for quaternary semiconductors from different experimental reports are rather scattered partially because of their structural complexity. The structural and chemical complexities of quaternary semiconductors also bring in additional computational and theoretical challenges. We are currently investigating the quasiparticle properties of other quaternary semiconductors and will present our results later.

## 5. SUMMARY

In summary, we have carried out a comparative study of the structural and electronic properties of Cu-based multinary semiconductors using first-principles electronic structure approaches. We point out that Cu *d* electrons in these systems have a dual nature. On one hand, being 3*d* electrons, they are intrinsically localized and experience a strong on-site Coulomb correlation. On the other hand, they are relatively shallow in energy and can hybridize with VI *p* electrons to form strong covalent bonds. An accurate account of both these aspects poses a significant challenge to theory. As a result, straightforward calculations within the LDA (or GGA) consistently underestimate the anion displacements. Anion displacement calculated using the HSE06 functional seems to improve considerably and better agrees with experiment. This is attributed to the fact that the HSE06 functional can better capture the short-range screened Hartree-Fock energy for localized *d* electrons.

Using the structures optimized within the HSE06 functional, we calculate the quasiparticle band structure of Cu-based multinary semiconductors using the many-body perturbation theory within the $G^0W^0$ approximation. The $G^0W^0$ approximation is carried out using the PBE, PBE+$U$, and HSE06 mean-field solutions as a starting point. Our results suggest that the PBE+$U$+$G^0W^0$ approach in general gives results that are better in agreement with experiments than those calculated with other approaches. Although the HSE06 functional gives an improved band gap and band structure near the band edge, and results in better anion displacements, it has certain limitations and does not seem to be able to correctly describe deep lying states. Using the PBE+$U$+$G^0W^0$ approach, we are able to reproduce the systematic trend of the chemical variation of the band gap of ternary semiconductors observed experimentally. Quasiparticle properties of two selected quaternary semiconductors (i.e., KS $Cu_2ZnSnS_4$ and KS $Cu_2ZnSnSe_4$) are also investigated. We predict that the band gap of $Cu_2ZnSnS_4$ is about 1.65 eV in the KS phase and 1.40 eV in the ST phase. These results are in consistent with available experimental values but require further verification. The band gap reduction in quaternary semiconductors as compared to their ternary analogs is attributed to the lowering of the anti-bonding CBM states. Furthermore, we have calculated and validated the value ($U$ = 4 eV) of the screened Coulomb energy for Cu $d$ electrons in these materials.

## ACKNOWLEDGMENTS

The authors would like to thank Dr. Shiyou Chen for useful discussions. This work is partially supported by National Basic Research Program (973-program) of China under Project No. 2007CB607503 and NSFC Grants (50825205, 50821004). P. Zhang is supported by the US National Science Foundation under Grant No. DMR-0946404. The calculations were carried out at Shanghai Supercomputer Center of China.

TABLE I. Calculated screened Coulomb and exchange matrix elements for the Cu $d$ electrons in CuGaS$_2$. The last line of each table shows the averaged value of the $U$ or $J$ matrix (see text for details).

| $U_{ij}$ (eV) | 1 ($e$ - 1) | 2 ($e$ - 2) | 3 ($t_2$ - 1) | 4 ($t_2$ - 2) | 5 ($t_2$ - 3) |
|---|---|---|---|---|---|
| 1 ($e$ - 1) | 6.21 | 4.21 | 4.88 | 4.88 | 4.18 |
| 2 ($e$ - 2) | 4.21 | 6.23 | 4.42 | 4.42 | 5.14 |
| 3 ($t_2$ - 1) | 4.88 | 4.42 | 6.11 | 4.38 | 4.38 |
| 4 ($t_2$ - 2) | 4.88 | 4.42 | 4.38 | 6.11 | 4.38 |
| 5 ($t_2$ - 3) | 4.18 | 5.14 | 4.38 | 4.38 | 6.12 |
| Avg. $U$ | | | 4.85 (eV) | | |

| $J_{ij}$ (eV) | 1 ($e$ - 1) | 2 ($e$ - 2) | 3 ($t_2$ - 1) | 4 ($t_2$ - 2) | 5 ($t_2$ - 3) |
|---|---|---|---|---|---|
| 1 ($e$ - 1) | 0 | 1.00 | 0.64 | 0.64 | 1.00 |
| 2 ($e$ - 2) | 1.00 | 0 | 0.88 | 0.88 | 0.52 |
| 3 ($t_2$ - 1) | 0.64 | 0.88 | 0 | 0.87 | 0.88 |
| 4 ($t_2$ - 2) | 0.64 | 0.88 | 0.87 | 0 | 0.88 |
| 5 ($t_2$ - 3) | 1.00 | 0.52 | 0.88 | 0.88 | 0 |
| Avg. $J$ | | | 0.82 (eV) | | |

TABLE II. Quasiparticle gaps of KS Cu$_2$ZnSnS$_4$ within the PBE+$U$ (+$G^0W^0$) approach applied to the $d$ electrons of different atoms. $\delta E_g$ is the difference of the calculated gaps within the PBE+$U$+$G^0W^0$ and the PBE+$G^0W^0$ approaches.

| $U$ [eV] | | | E$_g$ [eV] | | |
|---|---|---|---|---|---|
| Cu $3d$ | Zn $3d$ | Sn $4d$ | PBE+$G^0W^0$ | PBE+$U$+$G^0W^0$ | $\delta E_g$ |
| 4 | 0 | 0 | 1.077 | 1.572 | 0.495 |
| 0 | 4 | 0 | 1.077 | 1.107 | 0.030 |
| 0 | 0 | 4 | 1.077 | 1.082 | 0.006 |
| 4 | 4 | 4 | 1.077 | 1.645 | 0.568 |

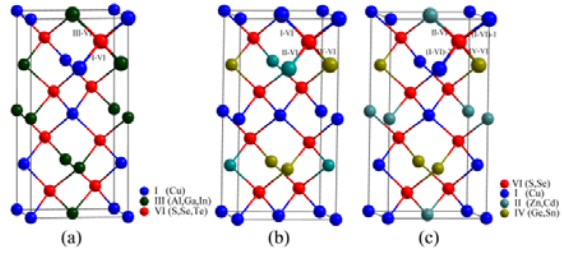

FIG. 1. (Color online) Crystal structures of (a) chalcopyrite (CH) ternary, (b) stannite (ST) quaternary, and (c) kesterite (KS) quaternary semiconductors.

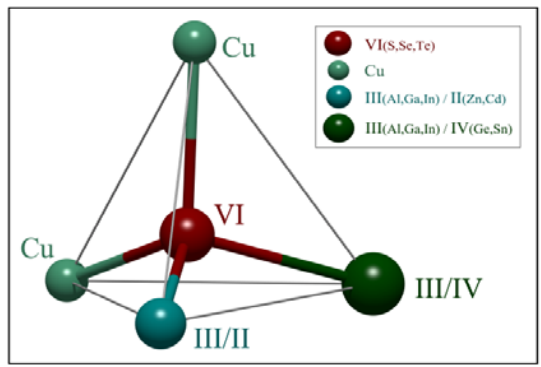

FIG. 2. (Color online) Building block of Cu-based ternary and quaternary semiconductors showing the displacement of the VI atom at the center.

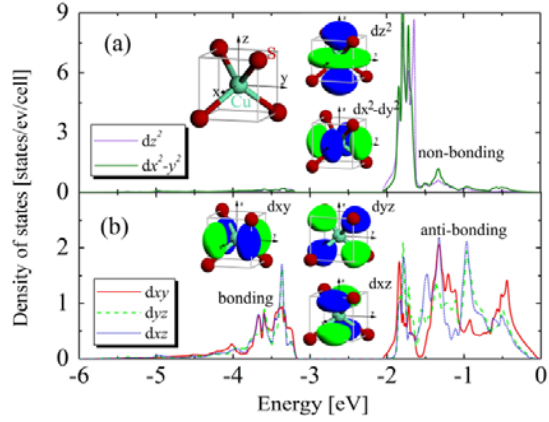

FIG. 3. (Color online) Density of states of CH CuGaS$_2$ projected on Cu $d$ states, (a) projected on $e$ orbitals and (b) $t_2$ orbitals. The Cu-centered cubic unit is shown in the insert. Orbital shapes of $e$ and $t_2$ in real space are also shown in the insert. The VBM is set to zero.

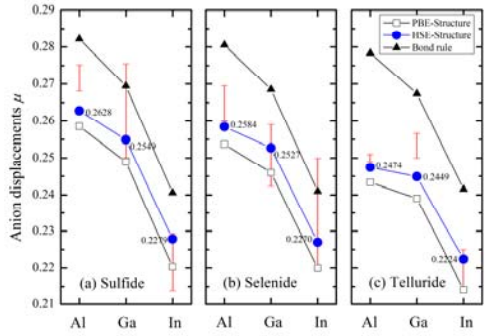

FIG. 4. (Color online) Anion displacements $\mu_{CH}$ of Cu-III-VI$_2$ (III=Al,Ga,In; VI=S,Se,Te) ternary semiconductors calculated within the PBE and the HSE06 functionals, as well as from the empirical *bond rule* (see text for details). The values calculated within the HSE06 functional are marked in the plot. Experimental data are marked with vertical bars.

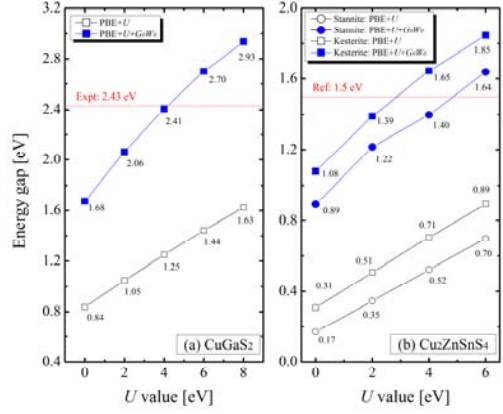

FIG. 5. (Color online) The $U$ dependence of band gaps of (a) CuGaS$_2$ (CH structure) and (b) Cu$_2$ZnSnS$_4$ (both the ST and KS structures) calculated within the PBE+$U$ and PBE+$U$+$G^0W^0$ approaches. The conventional PBE and PBE+$G^0W^0$ approaches are recovered when $U$ = 0 eV.

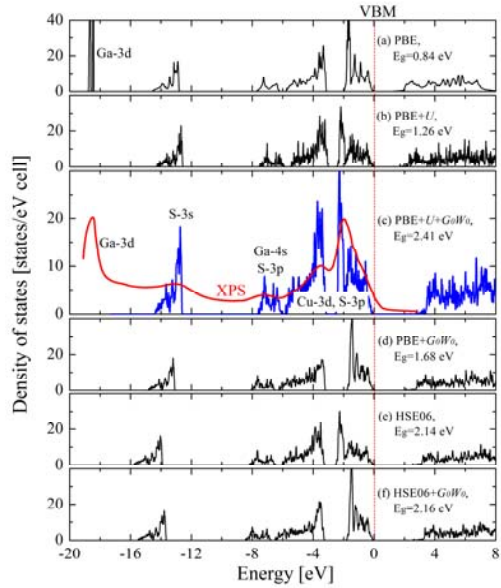

FIG. 6. (Color online) DOS of CH CuGaS$_2$ in the (a) PBE, (b) PBE+$U$, (c) PBE+$U$+$G^0W^0$, (d) PBE+$G^0W^0$, (e) HSE06, and (f) HSE06+$G^0W^0$ approaches. The XPS data are from Ref. [59] and superimposed on the PBE+$U$+$G^0W^0$ result. Ga 3$d$ electrons are taken as valence state in the panel of (a) and an on-site Coulomb energy of $U$ = 7 eV is applied on this state in order to reproduce the experimental result. The VBMs are set to zero.

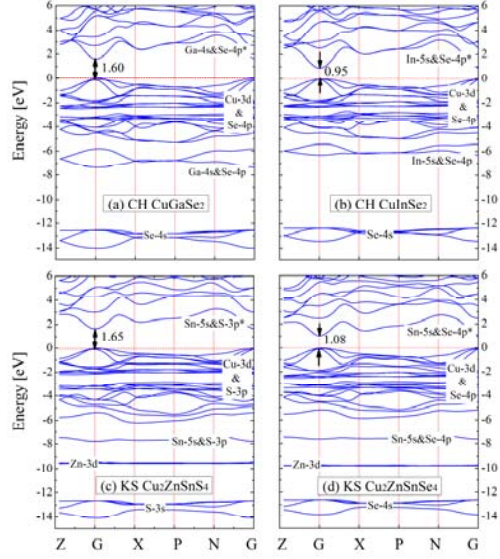

FIG. 7. (Color online) Quasiparticle band structures of (a) CH CuGaSe$_2$, (b) CH CuInSe$_2$, (c) KS Cu$_2$ZnSnS$_4$, and (d) KS Cu$_2$ZnSnSe$_4$ within the PBE+$U$+$G^0W^0$ approach. The VBMs are set to zero.

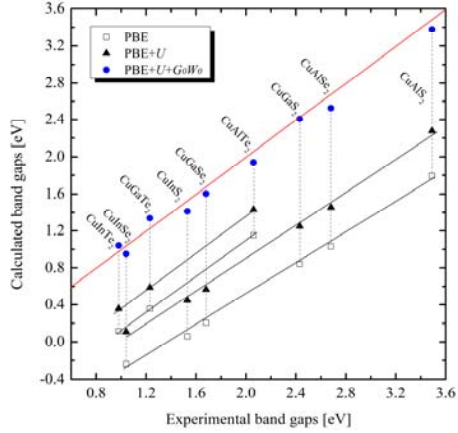

FIG. 8. (Color online) Band gaps of ternary semiconductors within the PBE, the PBE+$U$, and the PBE+$U$+$G^0W^0$ approaches. The experimental values (CuAlS$_2$, 3.49 eV;[59] CuGaS$_2$, 2.43 eV;[50] CuInS$_2$, 1.53 eV;[50] CuAlSe$_2$, 2.65 eV;[67] CuGaSe$_2$, 1.68 eV;[66] CuInSe$_2$, 1.04 eV;[68] CuAlTe$_2$, 2.06 eV;[44] CuGaTe$_2$, 1.23 eV;[69] CuInTe$_2$, 0.98 eV[70]) are taken for comparison.